\documentclass[twocolumn,prl,epsfig]{revtex4}
\usepackage{graphicx}

\begin{document}
\topmargin-1.0cm

\input epsf


\def\bea{\begin{eqnarray}}
\def\eea{\end{eqnarray}}
\def\ben{\begin{equation}}
\def\een{\end{equation}}
\def\benu{\begin{enumerate}}
\def\enu{\end{enumerate}}

\def\n{n}

\def\sss{\scriptscriptstyle\rm}

\def\g{_\gamma}

\def\l{^\lambda}
\def\lfc{^{\lambda=1}}
\def\lo{^{\lambda=0}}

\def\marnote#1{\marginpar{\tiny #1}}
\def\rsav{\langle r_s \rangle}
\def\invdif{\frac{1}{|\br_1 - \br_2|}}

\def\hatT{{\hat T}}
\def\hatV{{\hat V}}
\def\hatH{{\hat H}}
\def\1var{(\bx_1...\bx\N)}

\def\half{\frac{1}{2}}
\def\quart{\frac{1}{4}}

\def\bp{{\bf p}}
\def\br{{\bf r}}
\def\bR{{\bf R}}
\def\bu{{\bf u}}
\def\bmu{{\bf \mu}}
\def\bx{{\br t}}
\def\by{{y}}
\def\ba{{\bf a}}
\def\bq{{\bf q}}
\def\bj{{\bf j}}
\def\bX{{\bf X}}
\def\bF{{\bf F}}
\def\bchi{{\bf \chi}}
\def\bof{{\bf f}}

\def\cA{{\cal A}}
\def\cB{{\cal B}}

\def\x{_{\sss X}}
\def\c{_{\sss C}}
\def\s{_{\sss S}}
\def\xc{_{\sss XC}}
\def\Hx{_{\sss HX}}
\def\Hc{_{\sss Hc}}
\def\Hxc{_{\sss HXC}}
\def\xj{_{{\sss X},j}}
\def\xcj{_{{\sss XC},j}}
\def\N{_{\sss N}}
\def\H{_{\sss H}}

\def\ext{_{\rm ext}}
\def\pot{^{\rm pot}}
\def\hyb{^{\rm hyb}}
\def\HF{^{\rm HF}}
\def\hah{^{1/2\& 1/2}}
\def\loc{^{\rm loc}}
\def\LSD{^{\rm LSD}}
\def\LDA{^{\rm LDA}}
\def\GEA{^{\rm GEA}}
\def\GGA{^{\rm GGA}}
\def\SPL{^{\rm SPL}}
\def\sce{^{\rm SCE}}
\def\PBE{^{\rm PBE}}
\def\DFA{^{\rm DFA}}
\def\TF{^{\rm TF}}
\def\VW{^{\rm VW}}
\def\helm{^{\rm unamb}}
\def\una{^{\rm unamb}}
\def\ion{^{\rm ion}}
\def\HOMO{^{\rm HOMO}}
\def\gs{^{\rm gs}}
\def\dyn{^{\rm dyn}}
\def\adia{^{\rm adia}}
\def\I{^{\rm I}}
\def\pot{^{\rm pot}}
\def\sav{^{\rm sph. av.}}
\def\syv{^{\rm sys. av.}}
\def\pnav{^{\rm sym}}
\def\av#1{\langle #1 \rangle}
\def\unif{^{\rm unif}}
\def\LSD{^{\rm LSD}}
\def\ee{_{\rm ee}}
\def\vir{^{\rm vir}}
\def\ALDA{^{\rm ALDA}}
\def\VUC{^{\rm VUC}}
\def\PGG{^{\rm PGG}}
\def\GK{^{\rm GK}}
\def\atom{^{\rm atmiz}}
\def\trans{^{\rm trans}}
\def\SPA{^{\rm SPA}}
\def\SMA{^{\rm SMA}}

\def\sav{^{\rm sph. av.}}
\def\syv{^{\rm sys. av.}}

\def\up{_\alpha}
\def\dn{_\beta}
\def\up{_\uparrow}
\def\dn{_\downarrow}

\def\td{time-dependent~}
\def\KS{Kohn-Sham~}
\def\DFT{density functional theory~}

\def\fourint{ \int_{t_0}^{t_1} \! dt \int \! d^3r\ }
\def\fourintp{ \int_{t_0}^{t_1} \! dt' \int \! d^3r'\ }
\def\intx{\int\!d^4x}
\def\sph_int{ {\int d^3 r}}
\def\radint{ \int_0^\infty dr\ 4\pi r^2\ }

\def\PRA{Phys. Rev. A\ }
\def\PRB{Phys. Rev. B\ }
\def\PRL{Phys. Rev. Letts.\ }
\def\JCP{J. Chem. Phys.\ }
\def\JPCA{J. Phys. Chem. A\ }
\def\IJQC{Int. J. Quant. Chem.\ }
\input rotate
\def\epsH{\epsilon_{H}}
\def\epsL{\epsilon_{L}}
\def\tot{_{tot}}
\def\R{_{res}}

\author {Max Koentopp and Kieron Burke}
\affiliation {Department of Chemistry and Chemical Biology, Rutgers University, 610 Taylor Rd., 
Piscataway, NJ 08854}
\title{Zero-bias molecular electronics:
Exchange-correlation corrections to Landauer's formula}
\author{Ferdinand Evers}
\affiliation{Institute of Nanotechnology, Forschungszentrum Karlsruhe,
  76021 Karlsruhe, Germany
}
\date{\today}

\begin{abstract}
We show, that standard first principles calculations of transport
through single molecules miss exchange-correlation
corrections to the Landauer formula---the conductance is calculated at
the Hartree level. Furthermore, the
lack of derivative discontinuity in approximations can cause large
errors for molecules weakly coupled to the electrodes. 
From Kubo response theory,
both the Landauer formula and these corrections in the limit
of zero bias are derived and calculations presented.  
\end{abstract}

\maketitle

\newcommand{\upalda}{^{\scriptscriptstyle \rm ALDA}}
\newcommand{\downalda}{_{\scriptscriptstyle \rm ALDA}}
\newcommand{\vuc}{^{\scriptscriptstyle \rm VUC}}
\newcommand{\Tr}{{\rm Tr\; }}
\newcommand{\TrB}{{\rm Tr_{R}\; }}
\def\Trb#1{{\rm Tr\;}\left( #1 \right) }
\def\bE{{\bf E}}

Much experimental progress has been made in recent years in developing
methods to measure the conductance of single or few molecules in
between macroscopic leads,
and there is keen interest in the theoretical modeling of such
systems\cite{NR03}.
In the
case of organic molecules, covalently bound to the metallic
  electrodes, the transport properties are sensitive to the electronic structure,
  so chemical details are likely to be important,
and a first principles treatment is desirable. In order to describe
  the coupling of the molecule to the macroscopic leads in an
  appropriate way, parts of the leads must be included in the
  calculation.
  Given the number of atoms
required to simulate both the molecule and the first layers of the leads,
density functional theory (DFT) is an obvious choice.

Since the first sucessful conductance experiments for single molecules,
there have been several ground-breaking
calculations of this type, and a variety of codes perform
DFT-based calculations of I-V curves of single molecules between
metallic contacts\cite{BMOT02,DGD01,DPLb00,PDL02,HCWS02,EWK03b,XDR02}.  In these
calculations, a potential difference $V$ between the bulk electrodes is imposed.
A self-consistent {\em ground-state} Kohn-Sham (KS) calculation is performed
for the molecule plus a few layers of the leads.
Then, via Green's functions
the current is calculated using the celebrated two-terminal
Landauer formula\cite{L57}. The macroscopic leads 
enter via self--energies. We denote this ``standard'' approach by its common acronym,
NEGF. These calculations are parameter-free and 
often yield qualitative agreement with experiment, and so might appear to be as rigorous
as any DFT calculations. But detailed comparison for organic molecules
between Au-electrodes reveals quantitative discrepancies.  Conductances
are typically overestimated, often by one or two orders of magnitude\cite{EWK04}.

Neither the Hohenberg-Kohn theorem\cite{HK64},
which established ground-state DFT,
nor the 
Runge-Gross theorem for time-dependent problems\cite{RG84}, apply
to extended systems carrying current in homogeneous electric fields.
In consequence, questions have recently been raised about the validity
of the NEGF approach
\cite{RCBH03,EWK04}.  For example, the
calculated transmission is that of 
the KS potential.
In the case of a molecule weakly coupled to two leads, whose
KS levels are sharp, well-separated resonances, the NEGF
approach produces peaks in the conductance at the positions of
unoccupied levels of the KS system\cite{DPLb00}.
Such transitions are known to differ, in general, from the true excitations
of the interacting electronic system.

To tackle the transport problem rigorously for a finite bias is daunting,
and only recently have several suggestions been put 
forward\cite{DG04,BCG05,SAb04}.  
In the present paper, we examine only the weak bias regime,
i.e., the limit in which the potential difference across the
molecule is infinitesimal,
because here 
we can deduce the exact answer.

A primary result of this paper is to demonstrate rigorously that NEGF calculations
include only the Hartree response of the system.
It is alarming, that this level of calculation can be inadequate
already, when e.g. the respnse of an isolated molecule to a static
electric field is to be considered.
This lack is utterly independent of which standard approximate
functional is used:  All
DFT calculations to date suffer from this limitation.  It is inherent in the
methodology, just as for all Hartree calculations.
Second, we estimate the size of the XC corrections using 
the gradient expansion in the current of Vignale-Kohn\cite{VK96}.
Even when such contributions are small, the lack of
{\em derivative discontinuity}
in semilocal functional approximations for the
ground state likely produces significant errors.
Finally, we argue that, under certain conditions,
peak spacings in a zero-bias Coulomb
blockade experiment {\em are} accurately given by NEGF calculations.

Consider any system that can carry a DC current
in a specific direction (which we call the $z$ direction) and
that contains some atomic-sized barrier in this direction.
For simplicity, we analyze only the symmetric case.
We apply a weak uniform (also for simplicity) electric field in the $z$-direction, and use
time-dependent current density functional theory (TDCDFT) to
calculate the current response.

\begin{figure}
\unitlength1cm
\begin{picture}(8,6.5)
\put(-5.1,9){\makebox(12,6.5){
    \includegraphics{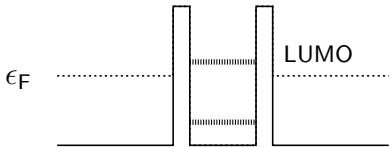}
}}
\put(0.9,6.1) {\large $\epsilon_{\sf F}$}
\put(4.6,6.4) {\sf LUMO}
\end{picture}
\vskip -5.3cm
\caption{Double barrier resonant tunnelling cartoon of a molecule
between two metallic leads; the lowest unoccupied molecular orbital
(LUMO) has been shifted and broadened relative to the isolated
molecule.}
\label{f:barr}
\end{figure}

The response of a system to a weak electric field is given by the Kubo
formula as
\ben
\bj(\br\omega) = \int d^3r'\; \hat \sigma (\br\br'\omega)\; \bE\ext (\br'\omega)\label{jtoE}
\een
where $\bj(\br\omega)$ is the first-order (physical) current density response to
the external electric field, and we use atomic units throughout.  Here
$\hat \sigma(\br\br'\omega)$ is the frequency-dependent non-local
conductance describing the response to the {\em external field},
rather than the 
response to the total electric field.

Within TDCDFT, the KS system is defined to reproduce the
time-dependent current density.  Thus
Eq. (\ref{jtoE}) becomes, for the KS system,
\ben
\bj(\br\omega) = \int d^3r'\; \hat \sigma\s[\n_0] (\br\br'\omega)\;
\left(\bE\tot(\br'\omega) + \bE\xc(\br'\omega)\right)
\label{jtoEs}
\een
where $\bj(\br\omega)$ is still the exact physical current response,
but now found from the KS non-local conductance (a functional of
the ground-state density $n_0$) applied
to the KS electric field, which includes both the total electric
field (external plus Hartree) 
and XC contributions.

The external field produces a finite
potential drop $V$ across  the barrier\cite{BG04}.
We restrict ourselves to one dimension, to demonstrate the principle.
In the limit in which $\omega\to 0$, the non-local conductance becomes
coordinate-independent \cite{BS89,KK01}.  Thus the left-hand side
becomes independent of $z$, and the integral over $z'$ applies
only to the potentials:
\ben
I = \sigma_{{\sss S}zz} (\omega=0) (V + V\xc)
\label{Land}
\een
where $V$ is the integral over the external and Hartree fields
and $V\xc$ is the induced net XC
potential drop in the vicinity of the barrier.
Equation (\ref{Land}), and the following interpretation, are the important results
of this Letter.  We analyze it in two steps.

{\em (i) Ignore $V\xc$}:
in the absence of the XC potential drop, Eq. (\ref{Land}) tells us
that the conductance, $I/V$, is just that of the {\em ground-state}
KS system. Careful derivations\cite{BS89} show that, for non-interacting systems, 
\ben
\sigma_{{\sss S}zz} (\omega=0) = T\s(\epsilon_F)/\pi
\een
where $T(\epsilon)$ is the transmission through all channels
through the barrier.
The resonances in the KS transmission function
translate into 
peaks in the conductance for the {\em interacting} system
{\em without} correction.

This brings us to the problem mentioned in the second paragraph,
namely the positions of the resonances in
the NEGF approach compared to the physical system.
Imagine the case of a one-dimensional double
barrier, as shown in fig.\ \ref{f:barr} as the ``molecule''.
Usually, $\epsilon_F$ is located in a spectral gap of the 
molecule,
(as in Fig. \ref{f:barr}),
so that the system is off-resonance and 
the conductance is non-vanishing only due to the
small overlap between the very weakly broadened levels
and $\epsilon_F$.

To probe the unoccupied resonances at zero bias, apply a gate voltage
$V_g$ to the molecule
perpendicular to the leads, shifting the LUMO down to $\epsilon_F$. 
As it passes through $\epsilon_F$ (as a function of $V_g$), there will
be a large peak in the conductance.
But consider what happens when the resonance begins to overlap with
$\epsilon_F$.
By virtue of its $\n_0$-dependence, 
the exact KS ground-state potential 
differs signficantly from the off-resonant case, altering the transmission
characteristics.  Peaks in transmission are {\em not} at the
position of $(V_g=0)-$unoccupied resonances.

For a sharp resonance, the transmission coefficient is given by
\ben
T(\epsilon)={(\frac{\gamma}{2})^2}/\left\{(\epsilon-\epsilon\R)^2+(\frac{\gamma}{2})^2\right\}
\label{T}
\een
where $\epsilon\R$ and $\gamma$ are the position and width
of the resonance which, in DFT, depend on the
partial occupation,
$0 \leq f \leq 1 $,
of the resonant level.
We will now see how the use of smooth, approximate density functionals influences the
  position and width  of the resonance.

 Using the spectral function $A(E)$  we can write expressions for the
  spectral density of states, $n(\epsilon)=\frac{1}{\pi}A$, as well
  as for the transmission $T=\frac{\gamma}{2}A$, to obtain a simple
  linear relationship between $n(\epsilon)$  and the transmission  of
  such a level, $n(\epsilon)=
  2\, T(\epsilon)/(\gamma\pi)$. The self-consistent $f$ is found from
  integrating over $n(\epsilon)$ as
\ben
f = \int_{-\infty}^{\epsilon_F} d\epsilon\ n(\epsilon)
= \half+\frac{1}{\pi}\tan^{-1} 
\left\{2 \frac{\epsilon_F-\epsilon\R(f)}{\gamma(f)}\right\}
\label{fdef}
\een
After inverting this,
\ben
T^{-1}(\epsilon_F) =
1 + \tan^2 \left\{\pi(f(\epsilon_F)-1/2)\right\}.
\label{Tf}
\een

In fig.\ 2, we plot the transmission over energy for
this situation, with the parameters given in the caption.
For this calculation we set the width constant ($\gamma(f)=\gamma_0$). The actual
dependence of $\gamma$ on $f$ is expected to be weak and
have little effect on transmission peaks.
Now$\Delta \epsilon=\epsH-\epsL$ is several eV, where
$\epsH$ is the highest occupied
orbital of the $N+1$-electron molecule, and $\epsL$ the
LUMO of the $N$-electron molecule.
 As in a NEGF
calculation using a semilocal functional,
$\epsilon\R$ always depends smoothly on $f$, and varies continuously
between $\epsL$ and $\epsH$, we obtain
(assuming $\epsilon\R=\epsL +f\Delta \epsilon$)
the dashed line.

For weakly coupled
leads, where $\gamma \ll \Delta\epsilon$ (at any occupation), the
Fermi level is pinned to the resonance ($\epsilon\R (f) \to
\epsilon_F$) for $f \neq 0$ or $1$, so $\epsilon_F=\epsL+\Delta
\epsilon f$ and
we obtain, using Eq.\ (\ref{Tf}) the dotted line in fig.\ 2.
Thus, in an NEGF calculation, Eq.\ (\ref{Tf}) always produces
a broad peak whose width is comparable to $\Delta \epsilon$.
For the case of a linear relation, the width is just $\Delta\epsilon/2$.

But this is entirely an artifact of smooth density functional approximations.
The real system has a sharp resonance centered
at $\epsH$.  The exact KS potential of the molecule jumps (relative
to the reservoir)
as soon as there is an infinitesimal occupation of the resonant level:
\ben
\epsilon_{res}=\epsL +\Theta(f + \eta)\Delta\epsilon,~~~~(\eta\to 0).
\een
Solving eq.\ \ref{fdef} for $\epsilon_F$, we obtain
a peak in the transmission of width $\gamma$ around $\epsH$---the
solid line in fig.\ 2.
This is the famous derivative 
discontinuity\cite{PPLB82} of DFT.

\begin{figure}
\unitlength1cm
\begin{picture}(8,6.5)
\put(-6.5,10){\makebox(12,6.5){
\includegraphics{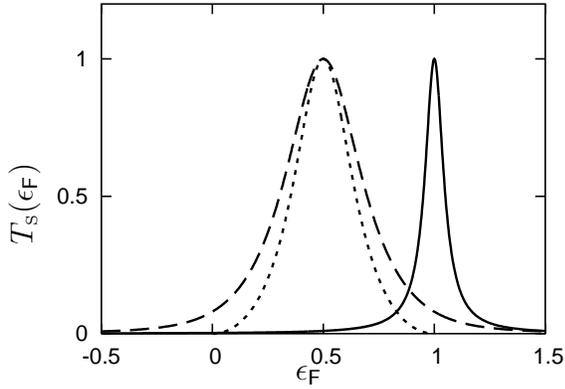}
}}
\setbox6=\hbox{\large $T\s(\epsilon_{\sf F})$}
\put(0.4,3.85) {\makebox(0,0){\rotl 6}}
\put(4,1.5) {\large $\epsilon_{\sf F}$}
\end{picture}
\vskip -1.5cm
\caption{Conductance peak from resonance: dashed line is self-consistent
approximate functional result, dotted line is approx result as $\gamma\to 0$,
and solid line is exact result.  Here $\epsL=0, \Delta\epsilon =1$ and
$\gamma_0=0.1$.}
\label{f:trans}
\end{figure}

Since the true transmission will be much more narrowly peaked than that in the
approximate DFT calculation, if the system is off-resonance, the DFT
calculation produces a strong overestimate of the true conductance.
Fig. \ref{f:trans} is a cartoon of this situation, in which the width of
the resonance is 10\% of the level shift, and a severe overestimate occurs
if the Fermi level is at, e.g., 0.5.

In reality, organic molecules may not be so weakly coupled to the
leads (although the widths of resonances in GGA
calculations are not a sure indicator of this, for reasons given above).
But this effect, or some remnant of it at less weak coupling,
would explain the
severe overestimate.
In Ref. \cite{EWK04}, the conductance of benzene was
calculated in two ways, a) by employing the standard approach based on the
KS energies and orbitals of an equilibrium DFT calculation (GGA)
and b) by replacing the KS data with their
counterparts obtained from a Hartree-Fock analysis. 
The typical transmission level in a window of 2$eV$ about
$\epsilon_F$ was reduced by an order of magnitude.
The best current way to see if this effect is the culprit for the overestimate
would be to perform exact exchange calculations\cite{KKP04},
which should contain most
of the effects of the true discontinuity.

{\em (ii) Include $V\xc$}:
Now we discuss how
to include:
\ben
V\xc= \int dz \ E_{z,{\sss XC}}(z,\omega\to 0)
\label{e9}
\een
where $\bE\xc$ is the XC electric field induced in response to the
applied field (and ignoring coupling between longitudinal and transverse
modes in $\sigma\s$).
We first note that, for any pure {\em density} functional,
$E\xc= -\nabla\cdot \delta v\xc$, so that $V\xc=\delta v\xc(z\to\infty)
-\delta v\xc(z\to
-\infty)$, i.e., the net induced XC potential drop from the extreme
left to the extreme right of the barrier.
In any semilocal
approximation, $V\xc$ therefore vanishes identically,
as the induced density response is localized to the region of the
barrier, so that far from the barrier, $\delta v\xc=0$.  Thus, using
common density functionals, the corrections to the KS Landauer formula
vanish!

The corrections are produced by non-local interactions present in the
exact XC functional. At least two mechanisms are
well established which generate such interaction terms 
non-local in the density. The most obvious one is exact exchange. 
Even a simple static exchange
calculation \cite{KKP04}, including response terms (i.e., beyond NEGF), might
yield a finite result, i.e., a net drop in the exchange potential
across the barrier, just as Hartree does.
A second mechanism is of the hydrodynamical type and therefore finds its
natural description within TDCDFT. For this case we are able to
offer an analytical estimate for the size of the effect. 
In the spirit of Ref. \cite{VUC97}, we use the Vignale-Kohn (VK)
approximation\cite{VK96} to obtain an expression for the purely viscous 
contribution to $V\xc$ (although the original derivation assumes
a high-frequency regime). 
 It involves a spatial variation of the density,
  $n(\bf x)$, which originates from the backscattering off the barrier.
  A rough estimate is obtained by assuming a) that 
  the most important variations in the density profile along the wire
  are of the Friedel-type and b) that the viscosity
  can be approximated by its static, homogenous value characteristic
  of a three-dimensional Fermi liquid. With these simplifications
  eq. (\ref{e9}) can be rewritten as 
\ben
V\xc/V \approx - (1-T(\epsilon_F))T(\epsilon_F)/(40 \pi^{1/2} k_F^{3/2}).
\een

The viscosity counteracts the current flow and reduces the
conductance. The factor $(1{-}T)$ takes into account that a barrier
causing reflection (and thus density inhomogeneities)
is needed for viscous flow to be generated. 

Since $k_F \approx 1$, the small prefactor (together with the fact
that $T{\leq}1$ for well resolved resonances) guarantees only small
corrections, as was found in a recent calculation\cite{SZVD05}.
This result, though suggestive, is not rigorous, however,
since it ignores both the
elastic hydrodynamic contribution and the limited validity of VK.

\begin{figure}[tb]
\begin{center}\leavevmode
\includegraphics[width=1\linewidth, angle=0]{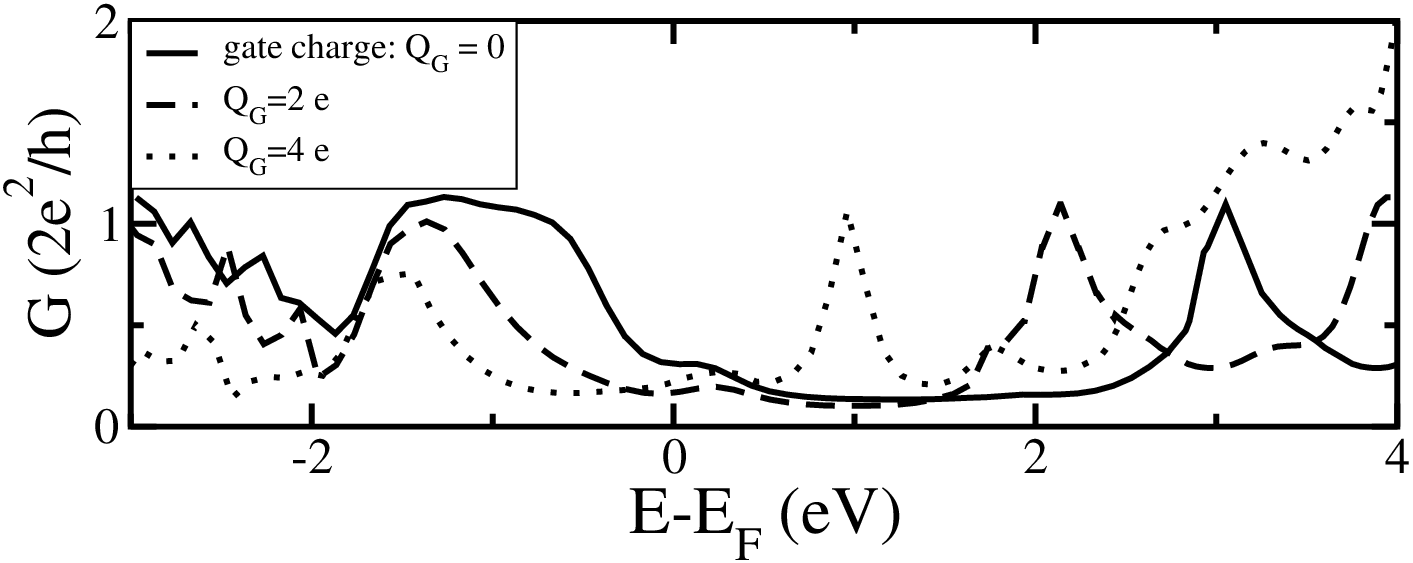}
\caption{Transmission of benzene-dithiolate in the presence of a
  gate. The gate is realized by a square sheet of homogenously
  distributed dummy charges (total sum is denoted by $Q_G$)
  located a distance of 3.8 \AA ngstrom above the carbon ring. }
\vskip -1cm
\label{f3}\end{center}\end{figure}

Finally, we show how, despite the fact that XC corrections to the
voltage, 
($V\xc$ in Eq. (\ref{Land}))
are missing, NEGF might be used to obtain
{\em exact} information to be compared with 
experiment.  Consider Coulomb blockade experiments, which measure zero-bias
conductance as a function of $V_g$ \cite{CB05}.
In Fig. \ref{f3} we 
show the transmission of a benzene ring, coupled via two sulphur
atoms to gold electrodes, obtained from a NEGF
transport calculation \cite{K05}.  The KS
LUMO moves towards $\epsilon_F$
with increasing gate charge (voltage). As argued above, the position of the LUMO
does not give a reliable estimate for the real peak
position. However, at the particular $V_g$ where the LUMO
passes through $\epsilon_F$, its energy must coincide with the real many-body level.
(In Fig. \ref{f3}, this would correspond to
$Q_G\approx 6$.) Therefore, at those $V_g$ where a KS-level
crosses $\epsilon_F$, a peak in the IV characteristics is
observed. The peak spacings
are given by a {\em ground-state} DFT calculation. 

Our calculation demonstrates the principle. 
It is missing the derivative
discontinuity, but  at least we can ignore the
missing $V\xc$.
There may be cases where the derivative discontinuity is
unimportant (i.e., for strong coupling or larger molecules) but
the missing $V\xc$ is not. Then standard NEGF calculations
will yield accurate results for peak spacings in Coulomb blockade
experiments, although the peak heights are strongly overestimated.

A not unlikely scenary, in which the $V\xc$ corrections are
especially large
is as follows.  Suppose that the quasistationary non-equilibrium state
with flowing current can be described by scattering from a single-particle
potential that differs from the ground-state KS potential.
Then the $V\xc$ as appearing in Eq. (\ref{Land})
must correct the KS off-resonant transmission sufficiently
to match that of the effective potential.

We conclude by noting that any formalism
to treat a many body problem that yields the Kubo response formula
when analyzed within TDCDFT will recover eq.\ (3)\cite{DG04,BCG05}. 
This work was supported by DOE under grant DE-FG02-01ER45928.
We thank Roberto Car, Achim Rosch, Peter W\"olfle and others for 
discussions.


\end{document}